\documentstyle[eqsecnum,aps]{revtex}

\begin{document}
\draft
\title{Axisymmetric Galaxy Distribution and the Center of the Universe}
\author{Yukio Tomozawa}
\address{The Michigan Center for Theoretical Physics and Randall Laboratory of\\
Physics, University of Michigan, Ann Arbor, MI. 48109-1120, USA }
\date{\today }
\maketitle

\begin{abstract}
The validity of Hubble's law defies the determination of the center of the
big bang expansion, even if it exists. Every point in the expanding universe
looks like the center from which the rest of the universe flies away. In
this article, the author shows that the distribution of apparently circular
galaxies is not uniform in the sky and that there exists a special direction
in the universe in our neighborhood. The data is consistent with the
assumption that the tidal force due to the mass distribution around the
universe center causes the deformation of galactic shapes depending on its
orientation and location relative to the center and our galaxy. The location
of the center is estimated to be at a distance $\ $\symbol{126}0.88/h Gpc in
the direction of l=$\ $135 $^{\text{o}}\pm $ 30$^{\text{ o}}$ and b = -35 $^{%
\text{o}}\pm $ 20 $^{\text{o}}$ in galactic coordinates ($\alpha $ = 01h
36m, $\delta $ = +26d 50m in equatorial J2000.0 or in the direction of the
Constellation Pisces. Further study of the deformation of galaxies such as
triaxial galaxies and non-axisymmetric spiral galaxies can be utilized for a
more accurate determination of the universe center.
\end{abstract}


Since the discovery of Hubble's law in 1929, and the big bang interpretation
of its data, the question lingers whether a center for the expansion exists
and if so where. The Hubble law, ${\bf v}=H_{0}{\bf r}$, yields the
relationship, ${\bf v}_{2}-{\bf v}_{1}=H_{0}({\bf r}_{2}-{\bf r}_{1})$for
any two galaxies with positions and velocities, ${\bf r}_{1}$, ${\bf v}_{1}$
and ${\bf r}_{2},{\bf v}_{2}$ respectively, where $H_{0}$ = 100 h km/s Mpc
is the Hubble constant (with h = 0.5 \symbol{126}0.85). The last equation
implies that every point appears to be the center of the expansion. Besides,
the distribution of galaxies surrounding us seems to be isotropic on some
large distance scale. Fig. 1 (a) shows the distribution of 23,011 bright
galaxies in galactic coordinates compiled in $\ $RC3\cite{rc3}. The galaxies
in the galactic plane are missing for the obvious reason that they are
prevented from observation due to our own Milky Way. Plots in equatorial
coordinates and supergalactic coordinates show a similar uniform
distribution. The numbers of galaxies on the two sides of the galactic plane
are 12,323 in the north and 10,628 in the south with a ratio of 1.1651. The
difference in these numbers may be partly due to a statistical fluctuation
and partly due to a historical bias in observation in the past. Anyway, any
study hereafter can be normalized with this ratio. Fig. 1 (b) shows the
velocity distribution of the compiled galaxies. The number of galaxies for
which the velocities, cz, are known is 10633 (5923 in the north and 4710 in
the south). The approximate upper bound is \symbol{126}15,000 km/s and the
average, \symbol{126}7,500 km/s, corresponds to a distance of 75h$^{-1}$ Mpc.

In Fig. 2 (a), (b) and (c), we show the distribution of galaxies with
circular shape in the galactic, supergalactic and equatorial J2000.0
coordinates, respectively. These galaxies correspond to those with the value
R25 = 0.00 (the error is in the range of \symbol{126}0.10) in RC3, where R25
stands for the logarithm of the ratio of the apparent major and minor axes.
The numbers of such galaxies are 682 in the north, 416 in the south with
1098 for the total. Also shown in Fig. 2 (d) is the velocity distribution,
which is similar to Fig 1 (b). The pertinent characteristics of Fig. 2 are
the following:

(i) Apparently circular galaxies are not distributed uniformly in the sky.
If the orientations of galaxies are distributed at random, Fig 2 (a)-(c)
should show statistically uniform distributions. It definitely shows the
directionality.

(ii) The southern part is more compact compared to the northern part. Since
a closed curve surrounding the pole is extended to all longitudes in Fig 1
(a), the views in other coordinates, supergalactic coordinates in Fig 2 (b)
and equatorial coordinates J2000.0 in Fig 2 (c), provide intuitively clearer
distribution plots. Fig 2 (b) shows similar shapes for both, northern and
southern, distributions. The normalized ratio of the northern to the
southern distribution is (682)/(416)/(1.1651) = 1.4071. The shapes in
supergalactic coordinates in Fig 2 (b) match this ratio. Note that the left
and right blobs in Fig 2 (b) approximately correspond to those in the north
and the south in Fig 2 (a).

Fig 3 (a)-(d) shows the distribution of galaxies with R25 = 0.01-0.10,
0.11-0.20, 0.21-0.40 and 0.41-1.00. They do not show a distinct
directionality except for a higher density in the neighborhood of (135d,
-35d). The total number of galaxies, the numbers in the north and south as
well as the normalized north/south ratio are given for sliced values of R25
in Table 1. In summary, a distinct directionality exists only for R25 =
0.00, and not for the other values of R25.

An important question is what is the meaning of this directionality in the
distribution of apparently circular galaxies. The author suggests that the
existence of a center of the universe is consistent with the presented data
and computes the distance and the direction to the center. He also suggests
a future study for improving the determination of these important
cosmological parameters.

(I) Galaxies are subject to tidal forces due to the gravity of the masses
which are contained in a sphere of radius equal to the distance from the
center to the galaxies in question. Circular galaxies in the radial
direction from the center through us stay circular in the presence of tidal
forces, while those in the direction perpendicular to the radial seen from
us are deformed by the squeeze and stretch of tidal forces and therefore
become elliptical. This explains the directionality in Fig. 2 (a)-(c).

(II) The direction to the center must be towards the south, since the
southern part in Fig. 2 (a) is more compact and less in number compared to
the northern part. The central value is in the neighborhood of l =$135%
{{}^\circ}%
$ and b = $-35%
{{}^\circ}%
$. The distance to the center, R, can be estimated from the relationship
(R+L)/(R-L) = $(1.4071)^{1/2}$, where L is the average distance to the
bright galaxies in RC3, which is assumed to be 75 h$^{-1}$ Mpc. Then we
obtain R = 11.74 L = $0.88h^{-1}$ Gpc. This is 1/3.4 of the Hubble distance, 
$cH_{0}^{-1}$, or equivalently z = 0.35. The absence of a node for each blob
in Fig 2 (d) is consistent with this solution, i.e., the center cannot be
inside the bright galaxies compiled in RC3. It is a remarkable coincidence
that the value of z = 0.35 is the point where departure from the Hubble law
starts to reveal itself\cite{sn}.

(III) The average positions of the distributions in the north and south in
Fig.2 (a) or equivalent ones in (b) and (c), are not diametrically opposite.
This implies that local fluctuation of galaxy distributions influences the
gravitational lines of force, so that the average positions are displaced
from a radial direction. In fact, there is up to a 16 percent of disparity
between the numbers of galaxies in the two hemispheres in the RC3
compilation, as mentioned earlier.

(IV) If the matter density is constant, it produces a gravitational force
that is linear in the radial direction, thus yielding a tidal force that is
spherically symmetric and attractive. Obviously, such a tidal force does not
explain the data of Fig.2 (a)-(c). Although the assumption of homogeneity
and isotropy in the Friedman-Robertson-Walker metric yields a constant
matter density at a fixed time, experimental evidence for those assumptions
is yet to come. As a matter of fact, the analysis of this letter finds that
the matter density of the universe is not constant. Finally, the author
suggests that the deformation of galaxies by the cosmic tidal forces which
was discussed in this letter should be taken into account for the systematic
study of triaxiality of galaxies\cite{galaxy} and asymmetric spiral galaxies
\cite{spiral1}\cite{spiral2}. From statistical analysis of these galaxy
deformations, one may be able to reach a more accurate determination of the
new cosmological parameter, the position of the center of the universe.

\bigskip 

The author would like to thank the members of the Physics Department and the
Astronomy Department of the University of Michigan for useful information.

\bigskip 

\bigskip 

Figure Caption.

Fig.1 Distribution of bright galaxies compiled in RC3$\cite{rc3}$. (a)
Galaxy distribution in galactic coordinates. (b) Velocity distribution vs.
galactic latitudes.

Fig.2 The distribution of apparently circular galaxies which have the value
R25= 0.00: (a) In galactic coordinates. (b) In supergalactic coordinates.
(c) In equatorial J2000.0 coordinates. (d) Velocity distribution vs.
galactic latitudes.

Fig.3 Distribution of galaxies in various ranges of the value of R25: (a)
0.01-0.10, (b) 0.11-0.20, (c) 0.21-0.40 and (d) 0.41-1.00.

Table Caption.

Table1. The total number of galaxies, the numbers in the north and south and
the normalized north/south ratio in various ranges of R25. 

\bigskip 

Table 1

\begin{tabular}{|l|l|l|l|l|}
\hline
Range of R25 &  Total &  North &  South &  North/South Normalized Ratio \\ 
\hline
\ \ \ \ \ 0.00-0.00 & 1098 & 682 & 416 &  \ \ \ \ \ \ \ \ \ \ \ \ 1.4071 \\ 
\hline
\ \ \ \ 0.01-0.10 & 4573 & 2330 & 2243 &  \ \ \ \ \ \ \ \ \ \ \ \ 0.8916 \\ 
\hline
\ \ \ \ 0.11-0.20 & 4448 & 2322 & 2126 &  \ \ \ \ \ \ \ \ \ \ \ \ 0.9374 \\ 
\hline
\ \ \ \ 0.21-0.30 & 3286 & 1714 & 1572 &  \ \ \ \ \ \ \ \ \ \ \ \ 0.9358 \\ 
\hline
\ \ \ \ 0.31-0.40 & 2246 & 1148 & 1098 &  \ \ \ \ \ \ \ \ \ \ \ \ 0.8974 \\ 
\hline
\ \ \ \ 0.41-o.50 & 1638 &  \ 861 &  \ 787 &  \ \ \ \ \ \ \ \ \ \ \ \ 0.9281
\\ \hline
\ \ \ \ 0.51-1.00 & 4045 & 2210 & 1835 &  \ \ \ \ \ \ \ \ \ \ \ \ 1.0337 \\ 
\hline
\end{tabular}


\begin{references}
\bibitem{rc3}  de Vaucouleurs, G. et. al., Third Reference Catalogue of
Bright Galaxies, Vol I-III (Spring-Verlag, New York, 1991).

\bibitem{sn}  Perlmutter, S. et. al., Discovery of a supernova explosion at
half the age of the Universe. Nature 391, 51-54 (1998).

\bibitem{galaxy}  Binney, J. \& Tremaine, S., Galactic Dynamics, (Princeton
University Press, Princeton,1987).

\bibitem{spiral1}  Schoenmakers, R. H. M., Franx, M. and de Zeeuw, P. T.,
Measuring non-axisymmetry in spiral galaxies. MNRAS 292, 349-364 (1997).

\bibitem{spiral2}  Schoenmakers, R. H. M., Asymmetries in Spiral Galaxies,
(University Press, Veenendaal, 2000) and references therein.
\end{references}
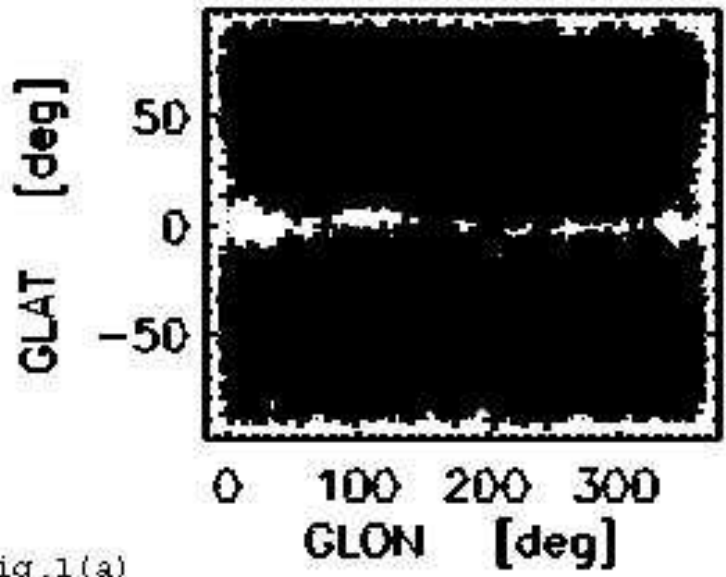
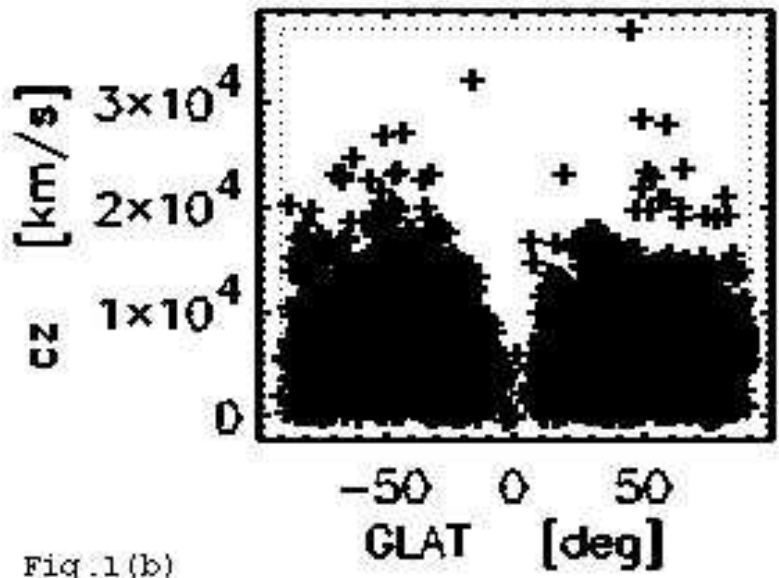
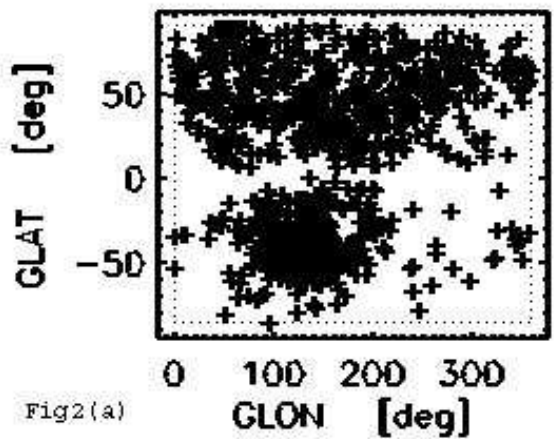
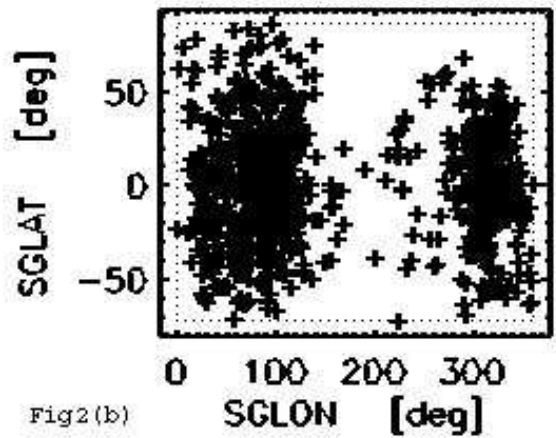
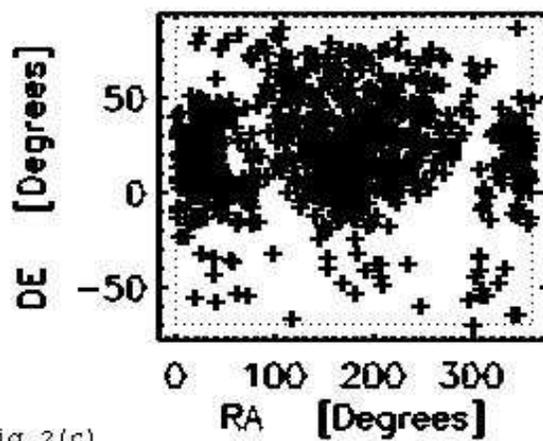
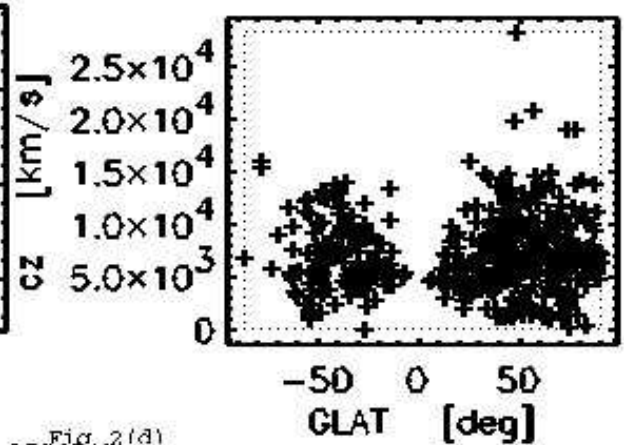
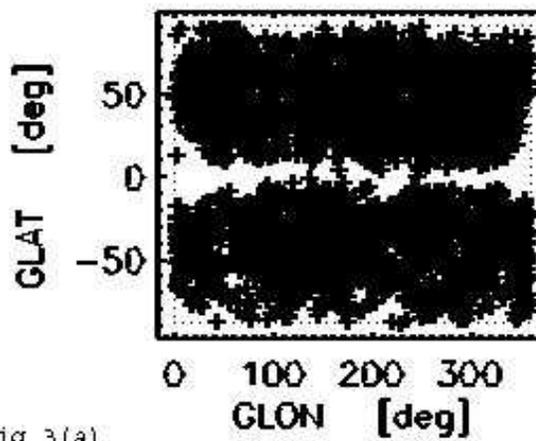
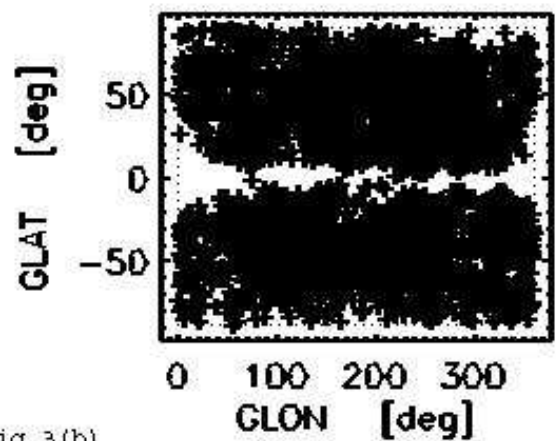
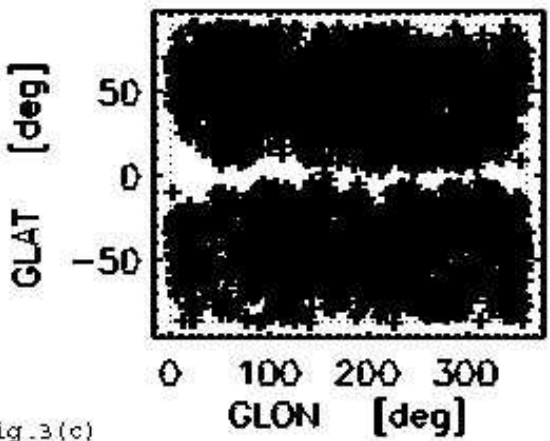
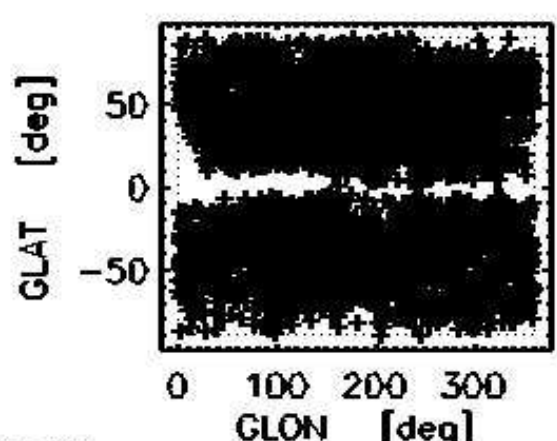
\end{document}